\documentstyle[amssymb,prb,aps,epsfig,multicol]{revtex}  
\begin{document}
\title{Electronic Structure of the Complex Hydride NaAlH$_4$}
\author{A. Aguayo$^1$ and D.J. Singh}
\address{Center for Computational Materials Science,
Naval Research Laboratory, Washington, DC 20375 \\
$^1$ also at School of Computational Sciences, George Mason University,
Fairfax, VA 22030}
\date{\today}
\maketitle

\begin{abstract}
Density functional calculations of the electronic structure of the
complex hydride NaAlH$_4$ and the reference systems NaH and AlH$_3$
are reported. We find a substantially ionic electronic structure for
NaAlH$_4$, which emphasizes the importance of solid state effects
in this material. The relaxed hydrogen positions in NaAlH$_4$ are
in good agreement with recent experiment. The electronic
structure of AlH$_3$ is also ionic.
Implications for the binding of complex hydrides
are discussed.
\end{abstract}


\begin{multicols}{2}

%\newpage
The complex hydrides, $A_x (M$H$_4)_y$, with $A$=Li, Na, K, Mg, Ca, Sr
or a mixture
of these and $M$=B or Al seem promising for H storage since they
contain very high weight percent H, and much of the H content can
be evolved at moderate temperatures.
\cite{bog1,gross,bog2,fichtner,zuttel}
However, hydrogen desorption
from these materials is associated with their melts, and they were long
viewed intrinsically non-cyclable. However, in 1997 Bogdanovic
and Schwickardi reported that with certain metal
additions, particularly Ti, NaAlH$_4$ can be cycled. \cite{bog1}
This discovery opened the door for consideration of these complex
hydrides as cyclable hydrogen storage materials. However,
this result has not been reproduced for other related complex hydrides
and understanding of the basic materials properties that govern
the uptake of H is still not complete. What is known is that hydrogen
desorption from NaAlH$_4$ takes place via a two step process, forming
first Na$_3$AlH$_6$, Al and hydrogen, and then NaH and Al metal.
Desorption is associated with the temperature where melting occurs.
While the Ti addition is often referred to as a catalyst or dopant,
the actual role of Ti in enabling the cyclability of NaAlH$_4$ has
yet to be established.
Possibilities include catalysis as mentioned,
modification of the thermodynamics of the decomposition,
{\it i.e.} the balance between solid state Ti containing NaAlH$_4$
and the decomposed NaH - Al - Ti mixture,
modification of the microstructure of post decomposition NaH + Al mixture,
{\it e.g.} by keeping the precipitated
Al and the NaH from segregating over large distances, {\it e.g.}
by enhancing the solubility of Al in NaH or
by the formation of Ti-Al alloys, like TiAl$_3$, instead of precipitated Al,
and others. 
\cite{gross,bog2,majzoub,bog3,bog4,sun,weidenthaler,balogh}

Developing understanding of the various phases involved will likely be
important for sorting out the physics of cyclable hydrogen storage
in this material. Here we start with the simplest of these phases,
NaAlH$_4$, by using density functional calculations of the electronic
structure, in comparison with results for NaH and AlH$_3$. Conventionally,
the bonding in NaAlH$_4$ is viewed as that of a salt made of Na$^{+}$
cations and AlH$_4^{-}$ anions, with the internal bonding of
the AlH$_4^{-}$ units being primarily covalent, consistent with
the tetrahedral coordination of Al and what is expected in the liquid.
We show that the electronic structure of solid NaAlH$_4$ is better
described as mixed ionic, {\it i.e.}, Na$^+$Al$^{3+}$H$_4^-$. Solid
NaH is found to be ionic as expected. AlH$_3$ is also ionic but with
a smaller band gap.
The ionic nature of solid NaAlH$_4$ is understood as
a result of long range Coulomb interactions, implying a greater sensitivity
of the electronic structure and therefore bonding of H
in NaAlH$_4$ to substitutions and
defects, than would otherwise be the case. Further this provides an
explanation for the association between the hydrogen desorption
and melting.

The present calculations were done within
the local density approximation (LDA)
to density functional theory,
using the general potential linearized augmented planewave method
with local orbital extensions, as implemented in the WIEN2K code.
\cite{singh-book,singh-lo,wien}
For consistency, the same LAPW sphere radii of 1.7 $a_0$ and 1.1 $a_0$
were used for the metal and hydrogen atoms, respectively,
in all three compounds. Well converged basis sets consisting of
an LAPW cutoff, $k_{max}$=5.91 $a_0^{-1}$ plus local orbitals were
used (the effective dimensionless values of the basis cutoff were
$R k_{max}=6.50$ for H and $R k_{max}=10.05$ for the metal atoms.
The Brillouin zone samplings were done using the special {\bf k}-points
method, with 21, 38 and 84 points in the irreducible wedge for
NaAlH$_4$, AlH$_3$ and NaH, respectively. This was found to be well
converged for these insulating materials.

\begin{figure}[tbp]
\centerline{\epsfig{file=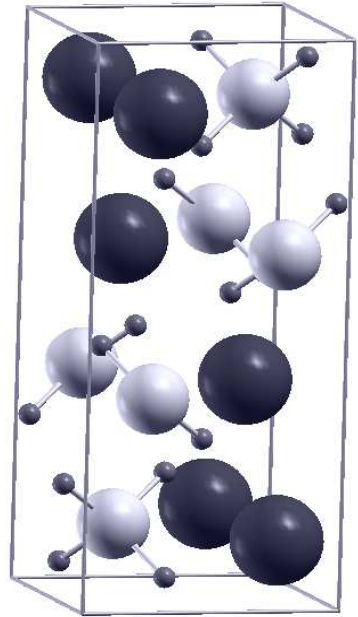,width=0.60\linewidth,clip=}}
\vspace{0.3cm}
\caption{Crystal structure of tetragonal NaAlH$_4$ with the
relaxed atomic positions. The small spheres are H, the large dark spheres
are Na and the large light spheres are Al.}
\label{naalh4-structure}
\end{figure}

NaAlH$_4$ occurs in a tetragonal structure (space group $I4_1/a$) with
lattice parameters $a$=5.021 \AA, $c$=11.346 \AA, and
its own structure type, which has two formula
units per primitive cell. \cite{belskii}
Recent neutron measurements for NaAlD$_4$ \cite{hauback} confirm this
structure, with slightly lower lattice parameters, but rather
different H positions. NaH occurs in the NaCl structure, with lattice
parameter $a$=4.88 \AA. \cite{zintl}
AlH$_3$ occurs in a rhombohedral structure (space group $R \bar{3} c$
or possibly $R \bar{3}$), with hexagonal lattice parameters,
$a$=4.451 \AA, $c$=11.766 \AA, and two formula units per primitive
rhombohedral cell
(from neutron and X-ray diffraction). \cite{turley,goncharenko,zogal}
In general, the structures
of hydrides, particularly the H positions,
are difficult to determine because
the H has a very small X-ray scattering factor and samples
can differ in H stoichiometry and ordering. Here we fully relaxed the
internal structures using LDA total energies and forces, keeping
the presumably reliable lattice parameters fixed at the reported
experimental values. For NaH there was no relaxation to do
because both atoms are on high symmetry sites. For NaAlH$_4$
we obtain positions differing from the positions reported
in Ref. \onlinecite{belskii}, but in close agreement with
very recent neutron results.
\cite{hauback}
The relaxed H positions (Wycoff notation, site 16$f$) are
$x$=0.2364, $y$=0.3906 and $z$=0.5451 as compared to
$x$=0.2371, $y$=0.3867 and $z$=0.5454 from neutron scattering.
This structure is illustrated in Fig. \ref{naalh4-structure},
which clearly shows the AlH$_4$ building blocks.
The Al-H bond length in our structure is 1.652 \AA.
This is only a little bigger than the sum of the covalent
radii of Al and H (0.37\AA + 1.18 \AA = 1.55\AA)
and would seem to be a reasonable number for covalently bonded AlH$_4^-$
units.
For AlH$_3$ we relaxed
in the lower symmetry $R \bar{3}$ spacegroup considered
by Zogal {\it et al.}. Our positions
differ somewhat from the X-ray structure of Ref. 
\onlinecite{zogal}, but are consistent with the assignment of
$R \bar{3}$ in the absence of H disorder.
The Al - H nearest neighbor distance in our structure
is 1.731 \AA, which is longer than in NaAlH$_4$ and is also longer
than the sum of the Al and H covalent radii. It should be mentioned that
this Al - H bond length is close to the value of 1.715 \AA, from
the structure of
Turley and Rinn. \cite{turley,alh3-struc}
We also considered $R3$, but obtained no further
relaxation (see below).

\begin{figure}[tbp]
\centerline{\epsfig{file=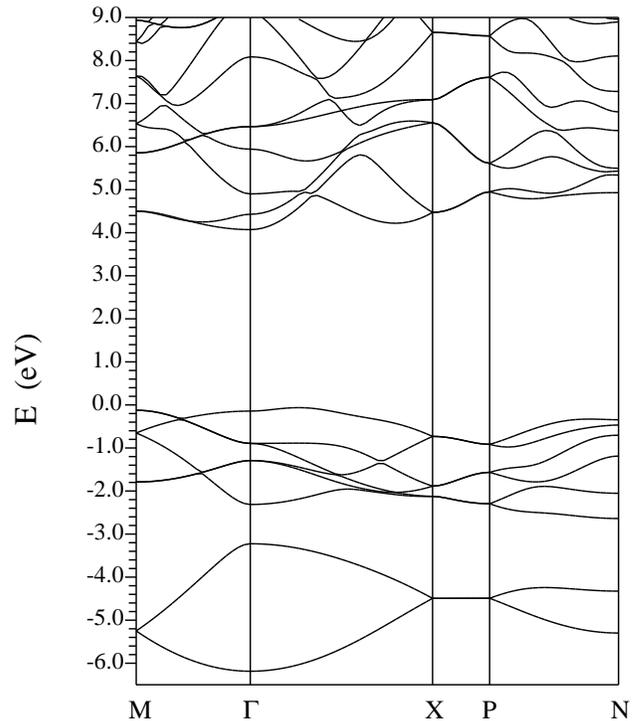,width=0.95\linewidth,clip=}}
\vspace{0.3cm}
\caption{LDA band structure of NaAlH$_4$ with the relaxed crystal structure.
The band gap is between a H derived valence band and metal derived
conduction bands.}
\label{naalh4-bands}
\end{figure}

\begin{figure}[tbp]
\centerline{\epsfig{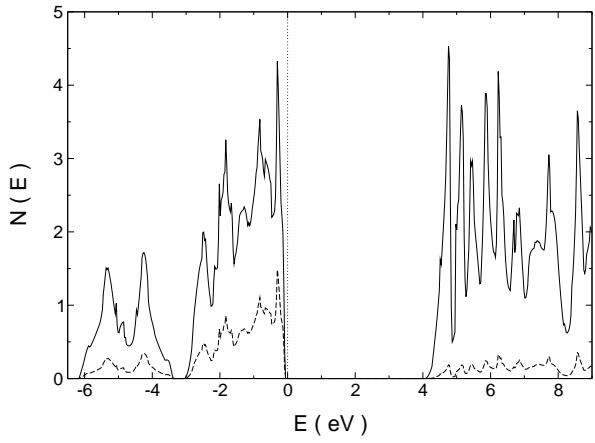}}
\vspace{0.3cm}
\caption{Electronic density of states and projection onto the H
LAPW spheres for NaAlH$_4$, on a per formula unit basis.
Note the ionic nature shown by
the very different hydrogen contributions to the valence and conduction bands.}
\label{naalh4-dos}
\end{figure}

\begin{figure}[tbp]
\centerline{\epsfig{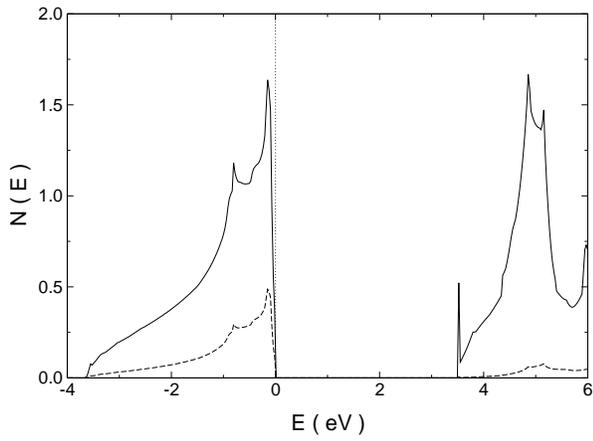}}
\vspace{0.3cm}
\caption{Electronic density of states and projection onto the H
LAPW sphere for NaH.}
\label{nah-dos}
\end{figure}

\begin{figure}[tbp]
\centerline{\epsfig{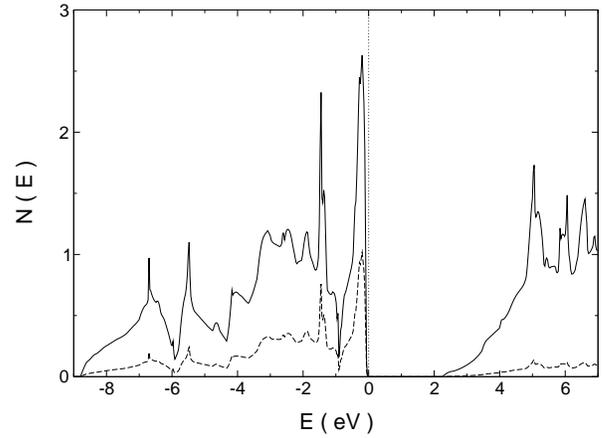}}
\vspace{0.3cm}
\caption{Electronic density of states and projection onto the H
LAPW spheres for AlH$_3$, on a per formula unit basis,
using the relaxed crystal structure.
Note
the smaller band gap but still ionic nature.}
\label{alh3-dos}
\end{figure}

The calculated LDA
band structure and corresponding electronic density of states (DOS)
for NaAlH$_4$ is shown in Figs. \ref{naalh4-bands}
and \ref{naalh4-dos}, respectively. The total DOS is similar
to that recently reported by Vajeeston and co-workers, \cite{vajeeston}
and is also similar to that reported
for the related compound LiAlH$_4$. \cite{vaj2}
The band structure has a large $\sim$4 eV band gap, separating
H derived valence bands from metal derived conduction bands.
We emphasize that despite the seeming AlH$_4$ units in the
structure, and the expected covalency of such chemical units,
the calculated electronic structure is very strongly ionic. In particular,
it can be seen from Fig. \ref{naalh4-dos} that the valence bands
are strongly dominated by H, while the conduction bands have very
much less H character.
The valence bands consist of two crystal field split manifolds,
each $\sim$3 eV in width.
The calculated DOS of NaH and AlH$_3$ are shown in Figs.
\ref{nah-dos} and \ref{alh3-dos}, respectively, along with the 
projections onto the H LAPW spheres. Since, the 1.1 $a_0$ spheres
are not large enough to fully contain the 1$s$ states of H$^-$ ions,
the H projection underestimates that H contribution to the electronic
structure. However, the ratio between the projections on H from different
energy regions, gives a good indication of the ratio of the H contributions
to the electronic structure in those energy regions.
The band structure of NaH is also strongly ionic, with a band
gap slightly smaller than that of NaAlH$_4$, while AlH$_3$ is also
ionic,
but has a smaller $\sim$2 eV band gap.
The valence band width of AlH$_3$ is $\sim$9 eV. It should be noted
that Goncharenko and co-workers \cite{goncharenko}, had
already conjectured that AlH$_3$ is ionically bonded based on
its crystal structure.

The reason for the ionic electronic structure of NaAlH$_4$
can be understood as due to the long range Coulomb interaction
in solids. This Ewald contribution to the energy favors
ionic electronic structures, and is well known to stabilize
O$^{2-}$ in metal oxides, for example, even though dimers and
small molecules with the same metal - O neighbors may be
covalent. Here H$^{-}$ is stabilized in this way.
In metal oxides, especially when there is some covalency between
the O and nominally unoccupied metal orbitals, the O$^{2-}$
ions are highly polarizable, as may be expected from the fact that
O$^{2-}$ outside the Coulomb field of the solid is not a stable ion.
Following the arguments of Cohen, \cite{cohen} which related
ferroelectricity in oxide perovskites to ionic electronic
structures with weak covalency of this type, and considering the electronic
structure of AlH$_3$, it seemed worthwhile to check if ferroelectricity
is present. Accordingly, we made small displacements of the atoms
away from the relaxed positions within the reduced symmetry
non-centrosymmetric spacegroup $R3$ and calculated the restoring
forces.
However, no ferroelectric instability was found
within this symmetry.

Returning to the bonding of NaAlH$_4$, we note some expected consequences
of the ionic electronic structure. First of all, since the bonding
is stabilized by long range interactions, rather than primarily
short range Al-H covalent bonds, it should be more sensitive to
stoichiometry, defects, lattice parameter changes and off-site substitutions,
than in a salt made of strongly covalent AlH$_4^-$ units.
This implies tunability of the hydrogen binding, {\it e.g.}
by alloying, which in turn would offer tunability of the thermodynamic
balance between the solid and the dehydrided NaH + Al mixture.
Secondly, it provides a natural explanation of why
the hydrogen desorption is strongly connected with melting. Presumably,
melting involves disruption of the H lattice and with it a loss of
the long range Coulomb stabilization of the H$^-$ ions. The resulting
loss of binding at melting then would result in H release from the material.
Finally, we note that while our calculations are specific to NaAlH$_4$,
the DOS of LiAlH$_4$ is qualitatively similar, \cite{vaj2}, suggesting
that similar physics may be operative there and in other related complex
hydrides.

We are grateful for helpful conversations with M. Gupta, R. Gupta,
G.-A. Nazri,
P. Vajda, Z. Wu, B. Yebka and K. Yvon.
Work at the Naval Research Laboratory is
supported by the Office of the Naval Research.

\end{multicols}

\begin{references}
\bibitem{alh3-struc}
In terms of rhombohedral lattice vectors, our relaxed structure
for AlH$_3$, has Al at (0,0,0) and (1/2,1/2,1/2) and H
at (0.751,0.389,0.111). The lattice vectors are ($\alpha,\beta,\beta$),
($\beta,\alpha,\beta$),($\beta,\beta,\alpha$), with $\alpha$=4.3626\AA,
and $\beta$=1.2153\AA.
\end{references}
\end{document}